\documentclass[12pt]{article}
\usepackage{amsmath,amssymb,epsfig}
\begin{document}
\parindent=0pt
\parskip=6pt
\rm

\vspace{0.5cm}

\begin{center}

{\bf \large Theoretical study of anisotropic layered antiferromagnets}

{\bf Dimo I. Uzunov}

G. Nadjakov Institute of Solid State Physics, Bulgarian Academy of Sciences,\\ Tsarigradsko Chaussee blvd. 72, Sofia, BG-1784 Bulgaria.

\end{center}

\begin{abstract}
We develop the field theory of antiferromagnets to layered structures on BCT crystal lattices with nearest-neibour and next-nearest-neighbour ferro- and/or antiferromagnetic interactions. For this aim the field theoretical counterpart of a lattice Heisenberg model is derived by standard theoretical methods: Hubbard-Stratonovich transformation and a generalized mean-field approach. We shown that the inter-layer interactions are a pure thermal fluctuation effect whereas the ground state is characterized by a perfect in-layer antiferromagnetic order and a lack of inter-layer coupling. This is a demonstration of 2D-3D dimensional crossover which is supposed to occur in real antiferromagnets, for example, in the spin-dimer antiferromagnet BaCuSi$_2$O$_6$.
\end{abstract}


\section[]{Introduction}

In this paper we consider (pseudo)spin-1/2 anisotropic antiferromagnets with body-centered tetragonal lattice (BCT) $(a,a,c)$ of volume $V=L^2L_z$, number of vertices $N = N_0^2N_z$, and lattice constants $(a,a,c)$. The spins $S(\boldmath$R$)$ are located at the vertices of the BCT lattice; $\boldmath
$R$ = (\boldmath$r$,z)$, including $\boldmath$r$ $ $= (x,y)$ and $z$, are the {\em discrete} coordinates of the spins.  Alternatively, these antiferromagnets can be defined by $N_z$ monolayers of $N^2_0$ spins $(S = \pm 1$) on the square ($x,y$), where the unit cell is given by($a,a$). The monolayers are at distance $c$ each other and the inter-layer exchange spin interactions ($J_{\perp}$) are either of antiferromagnetic type ($J_{\perp} < 0$), or, of ferromagnetic one ($J_{\perp} > 0$), whereas the in-layer spin exchange interactions are definitely of antiferromagnetic type ($J_{\parallel} < 0$). In all cases only nearest neighbor ($nn$) spin interactions are considered, which corresponds to the realistic point of view about the short-distance radius of the exchange interactions. Thus we have presented the three dimensional (3D) BCT lattice as interacting antiferromagnetic layers and this seems reasonable provided the inter-layer interactions are quite weaker than the in-layer ones ($|J_{\perp}| \ll |J_{\parallel}|$). Here we adopt this condition, although our study can be performed  beyond this restriction. Obviously, we shall consider a XXZ type of Heisenberg model of layered antiferromagnets.

Starting from a microscopic formulation of such type of systems we derive an effective field
Hamiltonian (generalized free energy \cite{ref01}) which describes the
thermodynamics, including relevant fluctuation phenomena in a
close vicinity of critical points of phase transitions. For this aim we apply a coarse-graining of the description [2,3]  in which the microscopic details of the particle behavior are smeared, if not absolutely ignored, but the quasi-macroscopic and true macroscopic (thermodynamic) behavior is correctly preserved in a conformity with the original microscopic model. Besides, following Refs. [4,5] we essentially use a more general framework of study. Our field theory might be used in interpretation of experiments and Monte Carlo simulations on real spin dimer systems, such as  BaCuSi$_2$O$_6$ with interesting antiferromagnetic properties (dimensional reduction at low temperatures, chirality and frustration effects)[6,7].

Usually, the studies of antiferromagnets are performed by dividing the original lattice in
two sublattices with magnetizations of opposite directions. Within
this scheme the actual order parameter is one of the sublattice
magnetizations, or, which is the same the half of the difference
between the two opposite magnetization vectors. Alternatively, for
a number of important problems, one may use the so-called
alternating magnetization. In both cases one should work with two lattice fields - the two sublattice magnetization fields, or, a linear transformation of the latter to other two lattice fields - the total and alternating magnetization fields. Such a two-field consideration meets some difficulties with the integral transformations (Hubbard-Stratonovich transformation, or, another suitable one [1,3,4])  but in our case the study can successfully be accomplished. Apart of the integral transformations of the spin variables the entire consideration includes several other (rotation, shift and Fourier)  transformations, application of the long wavelength approximation (LWLA), and for this reason the total derivation of the field theory is quite lengthy. In order to confine the paper in a manageable volume of several pages here we present a short outline of the derivation of the field theory (Sec. 2) with a brief discussion and enumeration of possible applications (Sec. 3).

\section[]{Derivation of a field theory}

We begin with the Heisenberg Hamiltonian:

\begin{equation}
\label{Eq1} \mathcal{H}=  \mathcal{H}_{\parallel}+\mathcal{H}_{\perp}+ \mathcal{H}_{h},
\end{equation}
where
\begin{equation}
\label{Eq2} \mathcal{H}_{\parallel}=-\frac{J_{\parallel}}{2}\sum_{\scriptsize
\mbox{\boldmath$r$},\mbox{\boldmath$z$},\mbox{\boldmath$\delta$}}
\mbox{\boldmath$S$}(\mbox{\boldmath$r$},\mbox{\boldmath$z$} ).\mbox{\boldmath$S$}(\mbox{\boldmath$r + \delta$},\mbox{\boldmath$z$})
\end{equation}
is the in-layer interaction term,
\begin{equation}
\label{Eq3} \mathcal{H}_{\perp}=-\frac{J_{\perp}}{2}\sum_{\scriptsize
\mbox{\boldmath$r$},z; \Delta\mbox{\boldmath$r$},\Delta z}
\mbox{\boldmath$S$}(\mbox{\boldmath$r$},z ).\mbox{\boldmath$S$}(\mbox{\boldmath$r$} + \Delta\mbox{\boldmath$r$}, z +\Delta z)
\end{equation}
is the inter-layer interaction term, and
\begin{equation}
\label{Eq4} \mathcal{H}_{h}=-\sum_{\scriptsize
\mbox{\boldmath$R$}}
\mbox{\boldmath$h$}(\mbox{\boldmath$R$}).\mbox{\boldmath$S$}(\mbox{\boldmath$R$})
\end{equation}
is the (Zeeman) energy of interaction with a non-uniform external magnetic field $H(\mbox{\boldmath$R$}) = h(\mbox{\boldmath$R$})/g\mu_B$; $g$ is the {\em g}-factor, and $\mu_B$ is the Bohr magneton. Usually the external field is uniform $h(\mbox{\boldmath$R$}) \equiv h$, but there are important cases when we should consider a space $(\mbox{\boldmath$R$}$-) dependent external filed \cite{ref01}. In Eq.~(2), the vector $\mbox{\boldmath$\delta$}$ runs values ($\pm a,0$) and   ($0,\pm a$), namely, describes the locations of the {\em nn} to any spin in a given layer, for example, the {\em z}th layer. In Eq.(3), which describes the energy of the inter-layer {\em nn} interactions, the vector $\Delta\mbox{\boldmath$r$}$ runs the vector values ($\pm a/2,0$) and   ($0,\pm a/2$), and the quantity $\Delta z$ takes values $\pm c/2$.

The Eqs. (1)-(4) represent the anisotropic antiferromagnet by the "lattice spin field" $S(\mbox{\boldmath$R$})$, which takes values at the vertices $\mbox{\boldmath$R$}$ of a BCT lattice with a unit cell ($a,a,c$). Our aim is to derive a Hamiltonian of a usual field theory, where the field depends on a spatial vector ($\mbox{\boldmath$R$} \equiv \mbox{\boldmath$x$} $) which takes any value in the volume $V$ of the system. Besides, the new field(s) take continuous values rather than discrete ones ($\pm 1$). Here we point out the main steps in the derivation of the field theory.

The first step is the separation of the BCT lattice in two sublattices (labeled by {\em a} and {\em b}) so that two neighbor layers belong to different sublattices. So, the body centered spins in the BCT lattice belong to the lattice $a$, and the other spins belong to the sublattice $b$, or, vice versa. The local sublattice spins are labeled by $a$ and $b$, i.e., $S_a(\mbox{\boldmath$R$})$ and $S_b(\mbox{\boldmath$R$})$; hereafter these labels will be used for other sublattice quantities.

 The next step is the integral transformation to fields which take continuous values. One may use suitable Hubbard-Stratonicich transformation [2-3], or, alternatively, a generalized self-consistent approach [4,5]. As shown for simpler cases, these two methods lead to one and the same result [4,5]. For our aims we  represent the total lattice field $S(\mbox{\boldmath$R$})$ by the sublattice fields $S_a(\mbox{\boldmath$R$})$ and $S_b(\mbox{\boldmath$R$})$ in the following way [2-3]
\begin{equation}
\label{Eq5}
S(\mbox{\boldmath$R$}) = \Delta_a(\mbox{\boldmath$R$}) S_a(\mbox{\boldmath$R$}) +\Delta_b(\mbox{\boldmath$R$}) S_b(\mbox{\boldmath$R$}),
\end{equation}
where the auxiliary fields $\Delta_a(\mbox{\boldmath$R$})$ and $\Delta_b(\mbox{\boldmath$R$})$ are given by
\begin{displaymath}
\label{Eq6}
\Delta_\rho = \left\{ \begin{array}{ll}
1, & \textrm{if $\boldmath$R$  \in \rho $}\\
0, & \textrm{otherwise},
\end{array} \right.
\end{displaymath}
where $\rho = a,b.$

Further, we represent the lattice field $S(\mbox{\boldmath$R$})$ as a sum of a statistical averaged value $\sigma(\mbox{\boldmath$R$}) = \langle S(\mbox{\boldmath$R$})\rangle$ and a fluctuation part $\delta S(\mbox{\boldmath$R$}) = S(\mbox{\boldmath$R$})- \sigma(\mbox{\boldmath$R$})$. The sublattice magnetizations $S_a(\mbox{\boldmath$R$})$ and $S_b(\mbox{\boldmath$R$})$ can also be represented in this way and then one defines the mean statistical fields $\sigma_a(\mbox{\boldmath$R$})$ and $\sigma_b(\mbox{\boldmath$R$})$. The standard (MF) treatment [1] implies a consideration of uniform statistical averages but here we keep the $\mbox{\boldmath$R$}$-dependence of all $\sigma$-fields.

The next step is the representation of the lattice Hamiltonian (1)-(4) by the the sublattice fields $\sigma_\rho(\mbox{\boldmath$R$})$ and the original lattice fields $S_\rho(\mbox{\boldmath$R$})$, ($\rho = a,b$), and accomplishment of the statistical summation over the latter. Solving the partition function with respect to the discrete fields $S_\rho(\mbox{\boldmath$R$}) = \pm 1$, one remains with a lattice field theory (generalized free energy, or, effective Hamiltonian) in terms of two continuous fields $\sigma_\rho(\mbox{\boldmath$R$})$, which may take any real value unless some special physical conditions introduce restrictions; but even in the latter case the allowed values of the fields $\sigma_\rho(\mbox{\boldmath$R$})$ remain continuous. Thus we make a progress although our $\sigma$-fields are still defined on the lattice and their spatial dependence remains discrete.

At this intermediate stage of consideration, the effective free energy (effective Hamiltonian) can be represented in the following way:
\begin{equation}
\label{Eq6} \mathcal{F} = \mathcal{H}^{(0)}
- \beta^{-1}\sum_{\scriptsize \mbox{\boldmath$r$}_{\alpha},z}\mbox{ln2ch}\beta|h^{(a)}_{\scriptsize\mbox{eff}}(\mbox{\boldmath$r$}_{\alpha}, z)| - \beta^{-1}\sum_{\scriptsize \mbox{\boldmath$r$}_{\beta},z}\mbox{ln2ch}\beta|h^{(b)}_{\scriptsize\mbox{eff}}(\mbox{\boldmath$r$}_{\beta}, z)|,
\end{equation}
where the subscripts $\alpha$ and $\beta$ run sites in the sublattices $a$ and $b$, respectively,  $\mathcal{H}^{(0)} = \mathcal{H}^{(0)}_{\parallel} +
\mathcal{H}^{(0)}_{\perp}$ with the Hamiltonian parts
\begin{equation}
\label{Eq7} \mathcal{H}^{(0)}_{\parallel}\! = \! \frac{J_{\parallel}}{2}\! \Big[\!\sum_{\scriptsize
\mbox{\boldmath$r$}_{\alpha},\mbox{\boldmath$\delta$}, z}\!\! \mbox{\boldmath$\sigma$}_b(\mbox{\boldmath$r$}_{\alpha}\! +\! \mbox{\boldmath$\delta$},z ).\mbox{\boldmath$\sigma$}_a(\mbox{\boldmath$r$}_{\alpha},z)\!  + \! \sum_{\scriptsize \mbox{\boldmath$r$}_{\beta},\mbox{\boldmath$\delta$}, z}
\mbox{\boldmath$\sigma$}_a(\mbox{\boldmath$r$}_{\beta}\! +\! \mbox{\boldmath$\delta$},z ).\mbox{\boldmath$\sigma$}_b(\mbox{\boldmath$r$}_{\beta},z)\Big]
\end{equation}
and
\begin{eqnarray}
\label{Eq8} \mathcal{H}^{(0)}_{\perp} \!\!&=&\!\! \frac{J_{\perp}}{2}\Big\{\!\! \sum_{\substack{\scriptsize
\mbox{\boldmath$r$}_{\alpha},\Delta\mbox{\boldmath$r$}\\ z, \Delta z }}\!\! \big[ \mbox{\boldmath$\sigma$}_a(\mbox{\boldmath$r$}_{\alpha}\! + \! \Delta\mbox{\boldmath$r$},z \! +\! \Delta z)\! + \! \mbox{\boldmath$\sigma$}_b(\mbox{\boldmath$r$}_{\alpha}\! + \! \Delta\mbox{\boldmath$r$},z\! + \! \Delta z) \big] .\mbox{\boldmath$\sigma$}_a(\mbox{\boldmath$R$}_{\alpha})  \nonumber \\
&& \!\! + \sum_{\substack{\scriptsize
\mbox{\boldmath$r$}_{\beta},\Delta\mbox{\boldmath$r$}\\ z, \Delta z }} \!\!\big[ \mbox{\boldmath$\sigma$}_b(\mbox{\boldmath$r$}_{\beta}\! + \! \Delta\mbox{\boldmath$r$},z \! +\! \Delta z)\! + \! \mbox{\boldmath$\sigma$}_a(\mbox{\boldmath$r$}_{\beta}\! + \! \Delta\mbox{\boldmath$r$},z\! + \! \Delta z) \big] .\mbox{\boldmath$\sigma$}_b(\mbox{\boldmath$R$}_{\beta})\Big\},
\end{eqnarray}
where $\mbox{\boldmath$R$}_{\alpha} \equiv (\mbox{\boldmath$r$}_{\alpha}, z)$ and $\mbox{\boldmath$R$}_{\beta}\equiv (\mbox{\boldmath$r$}_{\beta}, z)$. The Eq. (6) contains two effective fields which are given by
\begin{eqnarray}
\label{Eq9} h^{(a)}_{\scriptsize\mbox{eff}}(\mbox{\boldmath$r$}_{\alpha}, z) &=& h + J_{\parallel} \sum_{\scriptsize \mbox{\boldmath$\delta$}} \mbox{\boldmath$\sigma$}_b(\mbox{\boldmath$r$}_{\alpha} + \mbox{\boldmath$\delta$},z) \nonumber \\ && \!\!\!
+ J_{\perp}\!\!\sum_{\scriptsize \Delta\mbox{\boldmath$r$},\Delta z}\Big[  \mbox{\boldmath$\sigma$}_a(\mbox{\boldmath$r$}_{\alpha}\! +\!  \Delta\mbox{\boldmath$r$},z \! + \! \Delta z) + \! \mbox{\boldmath$\sigma$}_b(\mbox{\boldmath$r$}_{\alpha}\! + \!  \Delta\mbox{\boldmath$r$},z \! + \! \Delta z)  \Big]
\end{eqnarray}
and
\begin{eqnarray}
\label{Eq10} h^{(b)}_{\scriptsize\mbox{eff}}(\mbox{\boldmath$r$}_{\beta}, z) &=& h + J_{\parallel} \sum_{\scriptsize \mbox{\boldmath$\delta$}} \mbox{\boldmath$\sigma$}_a(\mbox{\boldmath$r$}_{\beta} + \mbox{\boldmath$\delta$},z) \nonumber \\ && \!\!\!\!\!
+ J_{\perp}\!\!\sum_{\scriptsize \Delta\mbox{\boldmath$r$},\Delta z}\!\!\Big[  \mbox{\boldmath$\sigma$}_b(\mbox{\boldmath$r$}_{\beta}\! +\!  \Delta\mbox{\boldmath$r$},z \! + \! \Delta z) + \! \mbox{\boldmath$\sigma$}_a(\mbox{\boldmath$r$}_{\beta}\! + \!  \Delta\mbox{\boldmath$r$},z \! + \! \Delta z)\!  \Big].
\end{eqnarray}

Further calculations are related with the Landau expansion [1-5] in power series of the fields, Fourier transformations of the fields, some auxiliary transformation of shift and rotation type, and LWLA. We denote the wave vectors by $\mbox{\boldmath$q$} = (\mbox{\boldmath$k$}, q_z)$, where $\mbox{\boldmath$k$} = (k_x,k_y)$. Within the LWLA, $k=|\mbox{\boldmath$k$}| \ll \pi/a$ and $q_z \ll \pi/c$, we shall keep only the fields with $q=|\mbox{\boldmath$q$}| \leq \Lambda$,  where the upper cutoff $\Lambda$ is supposed to be quite smaller than both $\pi/a$ and $\pi/c$ - the end points of the Brillouin zone. Thus, with the help of the LWLA, we perform a coarse-graining of the description by ignoring the fields with relatively large wave vector components, and keep only the fields with small momenta. The length $1/\Lambda$ should be of the order of the largest characteristic length of the system (interaction radii or correlation lengths far from critical points). In this way we neglect phenomena of microscopic size and ensure the description of quasi-macroscopic phenomena described with relatively large characteristic lengths up to the true macroscopic (thermodynamic) level.

The system sizes ($L, L_z$) are considered much larger than the length $1/\Lambda$, and this allows for the accomplishment of the continuum limit. The latter makes both wave vectors and spatial coordinates continuous variables. So, we achieve a usual field theory. In a consistency with the LWLA, only the first non-vanishing dependence of the self-energy Hamiltonian terms are taken into account. In our case, this dependence is of quadratic type ($\mbox{\boldmath$k$}^2, k_xk_y, q^2_z$), which corresponds to a gradient expansion to second order in the component of the nabla ($\nabla$-) operator. The wave vector dependence of the interaction terms - whose of fourth order in the fields, is ignored as irrelevant [1]. The more compact form of the final result for the generalized free energy (effective field Hamiltonian) is that in the space vectors $\mbox{\boldmath$R$}$ rather than in the $\mbox{\boldmath$q$}$-space representation of the fields.

Accomplishing the above mentioned steps, one obtains the following final result: $\mathcal{F} =\int d \mbox{\boldmath$R$}\hat{\mathcal{F}}(\mbox{\boldmath$R$})$, where the energy density $\hat{\mathcal{F}}$ is given by
\begin{eqnarray}
\label{Eq11}\hat{\mathcal{F}} &=& -2\beta\mu\mbox{\boldmath$h$}.\mbox{\boldmath$\sigma$} + \frac{1}{2}\big[r_f\mbox{\boldmath$\sigma$}^2 +c_f(\nabla_r\mbox{\boldmath$\sigma$})^2 + c_f^{\prime}(\nabla_z\mbox{\boldmath$\sigma$})^2 \nonumber \\ && + r_a\mbox{\boldmath$\varphi$}^2 +c_a(\nabla_r\mbox{\boldmath$\varphi$})^2 + c_a^{\prime}(\nabla_x\mbox{\boldmath$\varphi$}).(\nabla_y\mbox{\boldmath$\varphi$})    \big] \nonumber \\ &&
+ \frac{\beta^3}{6}\big\{ \mu^4\mbox{\boldmath$\sigma$}^4 +  \nu^4\mbox{\boldmath$\varphi$}^4 +2\mu^2\nu^2\big[ \mbox{\boldmath$\sigma$}^2\mbox{\boldmath$\varphi$}^2 + 2(\mbox{\boldmath$\sigma$}.\mbox{\boldmath$\varphi$})^2  \big] \big\}.
\end{eqnarray}
In Eq. (11), $\nabla_r = (\nabla_x,\nabla_y)$, and $\mbox{\boldmath$\phi$}$ and $\mbox{\boldmath$\varphi$}$ are three dimensional vector fields given by
\begin{align}
\label{Eq12}
\mbox{\boldmath$\sigma$} &= \mbox{\boldmath$\sigma$}_a + \mbox{\boldmath$\sigma$}_b + \frac{\mbox{\boldmath$h$}}{\mu},
\nonumber \\
\mbox{\boldmath$\varphi$}&= \mbox{\boldmath$\sigma$}_b - \mbox{\boldmath$\sigma$}_a.
\end{align}
The parameters of this theory are expressed by the microscopic parameters: $\mu = 2(J_{\parallel}  + 2J_{\perp})$, $\nu = 2J_{\parallel}$, and

\begin{align}
\label{Eq13}
r_f &= \mu \left(1-2\mu\beta \right),
\nonumber \\
c_f &= \left(J_{\parallel} +J_{\perp}\right)\left(2\mu\beta- \frac{1}{2}\right)a^2,\nonumber \\
c_f^{\prime} &= J_{\perp}\left(2\mu\beta- \frac{1}{2}\right)c^2,\nonumber \\
r_a &= -\nu\left(1+2\nu\beta\right),\nonumber \\
c_a &= \frac{\nu}{2}\left(2\nu\beta + \frac{1}{2}\right)a^2,\nonumber \\
c_a^{\prime} &= - 2J_{\perp}\left(2\nu\beta + \frac{1}{2}\right)a^2.
\end{align}
Obviously, the field $\mbox{\boldmath$\sigma$}$ is the magnetization in the presence of external magnetic field, whereas the field $\mbox{\boldmath$\varphi$}$ is proportional to the staggered magnetization - the antiferromagnetic order parameter field.

\section[]{Brief discussion}
Generally, the energy density (11) contains two order parameter fields - the magnetization field $\mbox{\boldmath$\sigma$}$ and the antiferromagnetic order parameter field$\mbox{\boldmath$\varphi$}$, and one may suppose that this theory may describe both ferromagnetic and antiferromagnetic phase transitions. However, as seen from the first equation (13), a condition for a spontaneous magnetization (at $\mbox{\boldmath$h$}= 0$) is given by the inequality $\mu >0$, i.e., $2J_{\perp} > |J_{\parallel}|$. For the case of weak inter-layer interaction $2J_{\perp} \ll |J_{\parallel}|$, a spontaneous magnetization has no chance to happen but antiferromagnetic structures are possible.

As seen from the gradient terms in Eq. (11), the antiferromagnetic fluctuations are of in-layer type, i.e., of 2D type. So, in a lack of interaction with the magnetization filed $\mbox{\boldmath$\sigma$}$, the antiferromagnet will behave as a 2D system. In case of effects of the magnetization fluctuations, including external field effects, the system may exhibit a 3D behavior. This is supported by the availability of the dependence of the magnetization fluctuations on the wave vector component $q_z$; see Eq. (11). While the theory contains thermal fluctuations, the fluctuation effects decrease with the decrease of the temperature. Thus one may conjecture that the ground state may exhibit a pure 2D behavior, while a 3D behavior is more probable at relatively high temperatures. The 2D-3D dimensional crossover in such antiferromagnetic systems has been discussed by the means of other methods [6,7] and the present theory could be used for a further elucidation of this topic.

In order to perform a more detailed study of the possible phases and their stability properties as well as of fluctuation phenomena and dimensional crossovers one should apply the MF approximation and filed-theoretical renormalization group techniques. The results can be used in interpretations of experiments and Monte Carlo simulations.

\end{document}